# KINEMATICS AND ELLIPSOIDAL MOTION OF THE MID TO LATE M-TYPE STARS

W. H. Elsanhoury[1,2] , M. I. Nouh[1] , Richard L. Branham Jr.[3], and Amnah S. Al-Johani[4]

[1]Astronomy Department, National Research Institute of Astronomy and Geophysics (NRIAG) 11421, Helwan, Cairo, Egypt. E-mails: welsanhoury@gmail.com & elsanhoury@nbu.edu.sa
[2]Physics Department, Faculty of Science and Arts, Northern Border University, Turaif Branch, Saudi Arabia.
[3]Instituto Argentino de Glaciología y Nivología (IANIGLA), C.C. 330, 5500, Mendoza, Argentina. E-mail: richardbranham1943@yahoo.com
[4]Mathematics Department, Faculty of Science, Tabuk University, Saudi Arabia, Email: xxwhitelinnetxx@hotmail.com & aalgohani@ut.edu.sa

**Abstract:** Truly, kinematics in the Solar neighborhood has provided important information both for the structure and for the evolution of the Galaxy since the early 20th century. The relationship between Oort constants and the ratio of the velocity dispersion are important quantities in stellar kinematics. In the present paper, we calculated the kinematical parameters and the Oort constants of various samples of late to intermediate M-type stars. We calculated the velocity dispersion $(\sigma_1, \sigma_2, \sigma_3)$ in units of km s$^{-1}$ for the samples under study. The longitude of the vertex $(l_2)$ having negative values with our analysis; i.e., the program I (538 stars), $l_2 = -0°.5410$, program II (100 stars), $l_2 = -0°.4937$ and program III (60 stars), $l_2 = -0°.9495$. We calculate the Oort constants as A = 14.69 ± 0.61 km s$^{-1}$ kpc$^{-1}$ and B = −16.70 ± 0.67 km s$^{-1}$ kpc$^{-1}$, and the rotational velocity $V_o = 257.38 \pm 9.40$ km s$^{-1}$. A possible explanation for the overestimated values of the second Oort constants has been presented.

## 1. INTRODUCTION

The cool dwarf stars, especially M-dwarfs (red dwarfs) are Galactic fossils with lifetimes much greater than the Hubble time (Laughlin & Adams 1997) and they are compelling targets for extrasolar planet surveys (Gaidos et al. 2007; Lunine et al. 2009; Charbonneau 2009), they are rare in the Solar neighborhood (Bochanski et al. 2013) and may have the greatest stellar component (∼ 70%) of the Galaxy (Kroupa et al. 2013). Comparing the number of single main-sequence stars with M$_V$ from 2.5 to 8.5 (corresponding to F, G, and K dwarfs) with the number with $M_V$ from 8.5 to 18.5 (corresponding to M-type dwarfs), one finds that F, G, and K dwarfs outnumber M-dwarfs in a magnitude-limited wide-field survey by more than 1000 to 1 (Muirhead et al. 2011).

Ultra-cool dwarfs (late M-dwarfs and brown dwarfs) with effective temperatures below 2,700 K make up a significant fraction of all stellar objects in the Galaxy (Martín et al. 1999; Kirkpatrick 2005).

Late M-dwarfs are originally discovered because they combined highly proper motion and a low luminosity (Kuiper 1939), they are low in masses ranging from (0.075 – 0.6 $M_\odot$), radii (0.08 – 0.62 $R_\odot$), temperatures (2,100 K – 3,800 K), faint in luminosity (5 mag fainter in $M_V$ than Sun-like stars, i.e., 0.001 – 0.08 $L_\odot$) (Kaltenegger & Traub 2009; Habson et al. 2018). Late type M dwarfs have an observational signature with a strong magnetic activity for a longer period of their lives compared to the Solar-mass stars (Delfosse et al. 1998; Mohanty & Basri 2003; Reiners & Basri 2007; West et al. 2015; Guinan et al. 2016). Because of their low mass and size, these enable detection and characterizing exoplanets via the transit and radial velocity techniques (Nutzman & Charbonneau 2008; Muirhead et al. 2011). The low temperature and high pressure in the photosphere of M-dwarfs raise severe problems for the computation of accurate atmosphere models (Chabrier & Baraffe 1997).

Main-sequence stars with masses below approximately (M < 0.35 $M_\odot$; mid-to-late M dwarfs) are fully convective depends on both ages and masses and are expected to have a different type of dynamo mechanism than Solar-type stars (Chabrier & Baraffe 1997; Newton et al. 2016). They have a spectral type ($\geq$ M3), with temperature ranging from 3,250 K to 2,300 K, radius (> 0.39 $R_\odot$), luminosity ($\leq$ 0.015 $L_\odot$) (Kaltenegger & Traub 2009), and they have similar kinematics as the inner halo and thick disk stellar populations (Kirkpatrick et al. 1991; Bochanski et al. 2013).

The present work aims to compute the ellipsoidal motion of mid-to-late M dwarfs and their kinematical parameters, which is divided into zones; Solar neighborhood ∼ 25 pc to ∼0.70 kpc, for different Galactic latitudes *(b)*; low and high latitudes, depending on the accuracy of available astrometric measurements, we have also derived their Galactic rotation constants.

The structure of the paper is as follows. In section 2 we described the samples of stars used. Section 3 is devoted to the kinematic model. The Oort constants are calculated in section 4. The discussion and conclusions are outlined in section 5.

## 2. SAMPLES OF MID-TO-LATE M DWARFS

The present analysis is based on two recent sources of data. The first one is from Newton et al. (2016) catalog, whose used photometry from the MEarth transit survey (Berta et al. 2012; Irwin et al. 2014) to identify rapid rotation period (and enhanced magnetic activity) for 1,883 mid-to-late M dwarfs in the northern hemisphere (MEarth-North), finding period ranged from 1 to 140 days through 2015.0 August 16. The vector components of the space motion (U, V, W; km s$^{-1}$) concerning the Local Standard of Rest (LSR) of 674 stars are computed and tabulated with a catalog via "The Strasbourg astronomical Data Center (CDS)[1]", which rely on the proper motions in combined with distance and radial velocities of these collected mid-to-late M dwarf data. The samples are divided into high Galactic latitudes $b \geq 20º$ and low one $b < 20º$, their information is:

i) Program I, 538 stars within Solar neighborhood $\approx$ 25 pc.
ii) Program II, 100 stars located above 25 pc till $\approx$ 50 pc.

The second source is drawn here with Zhang et al. (2019) following the surveys of cool subdwarfs from Data Release 4 (DR4; Gaia Collaboration et al. 2018) of the Large Sky Area Multi-Object Fiber Spectroscopic Telescope (LAMOST; Cui et al. 2012) spectra. In the Galactic anti-center direction with low Galactic latitudes, they finally identified M subdwarfs 2,791 stars whose spectral sequence ranged from M0 to M7 types. The data is available at http://paperdata.china-vo.org/szhang/DR4_Subdwarfs.csv[2] including a compilation of external data (photometric and Gaia Data Release 2 astrometric parameters), such that the radial velocities, proper motions, spectral types, and distances.

In the present work, we are setting Zhang et al. (2019) to download worksheet row data (program III) ranged from 25 pc to 0.70 kpc, considered the above attention with high Galactic latitudes $b \geq 20º$ and low one $b < 20º$. The adopted worksheet of 115-row data has been achieved. To improve our analysis, we used a recent and much larger data adopted today with the third intermediate Gaia Data Release (Gaia EDR3), hereafter EDR3[3] (Gaia Collaboration et al. 2020) collected by the European Space Agency's Gaia mission during 34 months of continuous scanning of the sky. EDR3 contains astrometry, i.e., central coordinates, proper motions, and parallaxes ($\alpha, \delta, \mu_\alpha \cos\delta, \mu_\delta, \pi$) for almost 2.5 billion (Torra et al. 2020) of stars and non-stellar sources, as well as three broad bands photometric magnitudes; G (330 – 1050 nm), the Blue Prism GBP (330 – 680 nm), Red Prism GRP (630 – 1050 nm) for sources brighter than 21 mag (Weiler 2018), and the heliocentric radial velocity ($V_r$; km s$^{-1}$). The quality of the Gaia astrometry is unprecedented: errors on the EDR3 measurements are of the order of 10 – 100 µas (20 – 130 µas on parallaxes and 20 – 140 µas yr$^{-1}$ on proper motions, Lindegren et al. 2021).

---

[1] http://vizier.u-strasbg.fr/viz-bin/VizieR?-out.add=_r&-source=J%2FApJ%2F821%2F93
[2] http://vizier.u-strasbg.fr/viz-bin/VizieR?-out.add=_r&-source=J%2FApJS%2F240%2F31

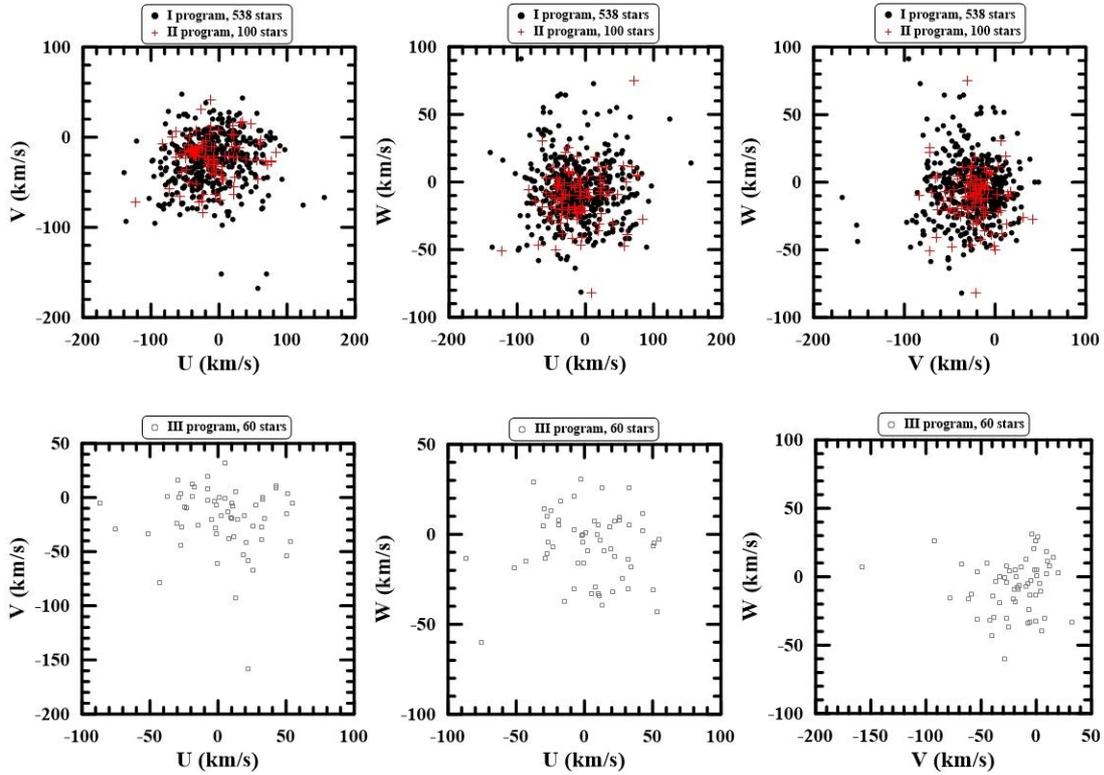

**FIGURE 1:** Plots of space velocity components with (V *vs.* U), (W *vs.* U), and (W *vs.* V) concerning the LSR of the mid-to-late M dwarfs. The *upper panel* adopted with Newton et al. (2016), which is corresponding to program I (538 stars, black closed circles) within the Solar neighborhood ≈ 25 pc and program II (100 stars, red plus symbols) located above 25 pc till ≈ 50 pc. The *lower panel* is concerned with program III (60 stars, gray open circles) located above 25 pc till ≈ 0.70 kpc adopted with Zhang et al. (2019).

---

[3] http://vizier.u-strasbg.fr/viz-bin/VizieR-3?-source=I/350&-out.add=_r

Next, we used the list of the row data devoted with Zhang et al. (2019) and interfere with those with EDR3 via crossmatch to get a corresponding row data devoted with EDR3 utilizing TOPCAT[4] (Taylor et al. 2005) based on The Star link Tables Infrastructure Library (STIL). As a result of these above procedures, we have 60 points (weighted statistically). Figure 1 shows the distribution of the vector space velocities $(U, V, W; \text{ km s}^{-1})$ (see section 3) of the three program stars.

## 3. THE VELOCITY ELLIPSOID PARAMETERS (VEPs)

The kinematic properties of any group of stars reflect its dynamical history and, by inference, the dynamical characteristics and evolution of the Galaxy itself (Mihalas & Binney 1968). For the stellar associations whose located near the Galactic plane, the primary results that emerge from an examination of the peculiar velocities of these disk stars relative to LSR reveal that they have very nearly a random (Gaussian) distribution in all three residual-velocity components $(U, V, W; \text{ km s}^{-1})$. An extremely elegant description of this situation in terms of a velocity ellipsoid was proposed here (Mihalas & Binney 1981).

The problem could be formulated as follows, suppose that we have the distribution of the residual velocities of these $(N_i)$ stars, coordinated with $(\alpha, \delta)$ located at a distance $(d_i; \text{pc})$, with proper motions in both directions $(\mu_\alpha \cos\delta, \mu_\delta; \text{ mas yr}^{-1})$, and heliocentric radial velocities $(V_r; \text{ km s}^{-1})$ and we would compute the velocity ellipsoids and their corresponding parameters, e.g., matrix elements $(\mu_{ij})$, space internal motions or called the velocity dispersion $(\sigma_1, \sigma_2, \sigma_3; \text{ km s}^{-1})$, direction cosines $(l_j, m_j, n_j; \forall_j = 1,2,3)$, the Galactic longitude of the vertex $(l_2)$, and determination of the Solar motion. In the following points, we will briefly present the basics formulae used in the algorithm.

i) The space velocity components $(U, V, W; \text{ km s}^{-1})$ as a function of the space stellar velocities $(V_x, V_y, V_z; \text{ km s}^{-1})$ could be written as (Smart 1968 and Liu et al. 2011).

$$U = -0.0518807421\, V_x - 0.872222642\, V_y - 0.4863497200\, V_z, \quad (1)$$

$$V = +0.4846922369\, V_x - 0.4477920852\, V_y + 0.5713692061 V_z, \quad (2)$$

$$W = -0.8731447899\, V_x - 0.1967483417\, V_y + 0.4459913295\, V_z. \quad (3)$$

and the mean velocities are given by.

---
[4] http://www.star.bris.ac.uk/~mbt/topcat/

$$\bar{U} = \frac{1}{N}\sum_{i=1}^{N} U_i, \qquad \bar{V} = \frac{1}{N}\sum_{i=1}^{N} V_i, \qquad \bar{W} = \frac{1}{N}\sum_{i=1}^{N} W_i.$$

where

$$V_X = -4.74 d_i \mu_\alpha \cos\delta \sin\alpha - 4.74 d_i \mu_\delta \sin\delta \cos\alpha + V_r \cos\delta \cos\alpha, \tag{4}$$

$$V_Y = +4.74 d_i \mu_\alpha \cos\delta \cos\alpha - 4.74 d_i \mu_\delta \sin\delta \sin\alpha + V_r \cos\delta \sin\alpha, \tag{5}$$

$$V_Z = +4.74 d_i \mu_\delta \cos\delta + V_r \sin\delta. \tag{6}$$

ii) Compute the $(3 \times 3)$ matrix elements of $(\mu_{ij})$ that controls the eigenvalue problem $[D(\lambda) = |B - \lambda I| = 0]$ for the velocity ellipsoid, i.e.

$$\mu_{11} = \frac{1}{N}\sum_{i=1}^{N} U_i^2 - (\bar{U})^2; \quad \mu_{12} = \frac{1}{N}\sum_{i=1}^{N} U_i V_i - \bar{U}\bar{V};$$

$$\mu_{13} = \frac{1}{N}\sum_{i=1}^{N} U_i W_i - \bar{U}\bar{W}; \quad \mu_{22} = \frac{1}{N}\sum_{i=1}^{N} V_i^2 - (\bar{V})^2; \tag{7}$$

$$\mu_{31} = \frac{1}{N}\sum_{i=1}^{N} V_i W_i - \bar{V}\bar{W}; \quad \mu_{33} = \frac{1}{N}\sum_{i=1}^{N} W_i^2 - (\bar{W})^2;$$

where, (N) is the number of member stars, ($\lambda$) is the eigenvalue and (B) could be written as; $B = \begin{vmatrix} \mu_{11} & \mu_{12} & \mu_{13} \\ \mu_{12} & \mu_{22} & \mu_{23} \\ \mu_{13} & \mu_{23} & \mu_{33} \end{vmatrix}$.

iii) The direction cosines $(l_j, m_j, n_j; \forall j = 1,2,3)$, the eigenvalue problem $(\lambda_j)$, matrix elements $(\mu_{ij})$, and (internal) velocity dispersion $(\sigma_j)$ [i.e. $\sigma_j = \sqrt{\lambda_j}; \forall j = 1,2,3$] along three axes is mathematically given by the following:

$$l_j = [\mu_{22}\mu_{33} - \sigma_j^2(\mu_{22} + \mu_{33} - \sigma_j^2) - \mu_{23}^2]/D_j, \tag{8}$$

$$m_j = [\mu_{23}\mu_{13} - \mu_{12}\mu_{33} + \sigma_j^2 \mu_{12}]/D_j, \tag{9}$$

$$n_j = [\mu_{12}\mu_{23} - \mu_{13}\mu_{22} + \sigma_j^2 \mu_{13}]/D_j, \tag{10}$$

where $(l_j^2 + m_j^2 + n_j^2 = 1)$ as an initial test for our code and,

$$D_j^2 = (\mu_{22}\mu_{33} - \mu_{23}^2)^2 + (\mu_{23}\mu_{13} - \mu_{12}\mu_{33})^2 + (\mu_{12}\mu_{23} - \mu_{13}\mu_{22})^2 + 2[(\mu_{22} + \mu_{33})(\mu_{23}^2 + \mu_{22}\mu_{33}) + \mu_{12}(\mu_{23}\mu_{13} - \mu_{12}\mu_{33}) + \mu_{13}(\mu_{12}\mu_{23} - \mu_{13}\mu_{22})]\sigma_j^2 + (\mu_{33}^2 + 4\mu_{22}\mu_{33} + \mu_{22}^2 - 2\mu_{23}^2 + \mu_{12}^2 + \mu_{13}^2)\sigma_j^4 - 2(\mu_{22} + \mu_{33})\sigma_j^6 + \sigma_j^8.$$

iv) The determination of the Solar motion concerning stars in their neighborhood is a necessary first step for investigations of stellar kinematics (Jaschek & Valbousquet 1991). Traditionally, three somewhat different methods can be employed for Solar motion determination (Jaschek & Valbousquet 1991, 1992, 1993, and 1994), using radial velocity data alone, proper motion data alone, or space motions (both radial and transverse components). The first two methods can be used for stars whose distances are not known (Mihalas & Binney 1981).

In carrying out these analyses, it is essential to restrict attention to homogeneous groups of stars. In our work, we consider the first and third procedures, to determine the Solar motion concerning the three program stars. This choice is due to the reason that; the data needed for these procedures are available, furthermore; proper motion alone permits only to determine the direction of the Solar velocity, but not its modulus. In what follows, we will be carried throughout the following points:

### a) The solution from space velocities

Knowing the distances of stars from direct parallax measurements, we can convert their observed proper motions to transverse velocities. Combining the transverse components with radial velocities, we can compute the space motions of the stars relative to the Sun. Velocity components in the equatorial system can be found directly from Equations of $(V_x, V_y, V_z;\ km\ s^{-1})$; and these components can then be expressed in the Galactic coordinate system by a simple rotation of axes, which yields $(U, V, W;\ km\ s^{-1})$ as defined in Equations (1-3).

Consider a group with spatial velocities $(\overline{U}, \overline{V}, \overline{W};\ km\ s^{-1})$. The components of the Sun's velocities are $(U_\odot, V_\odot,\ and\ W_\odot)$ are given as; $(U_\odot = -\overline{U})$, $(V_\odot = -\overline{V})$, and $(W_\odot = -\overline{W})$. Therefore, we have the Solar elements with spatial velocities considered like.

$$S_\odot = \sqrt{\overline{U}^2 + \overline{V}^2 + \overline{W}^2}, \tag{11}$$

$$l_A = \tan^{-1}\left(\frac{-\overline{V}}{\overline{U}}\right), \tag{12}$$

$$b_A = \sin^{-1}\left(\frac{-\overline{W}}{S_\odot}\right). \tag{13}$$

where $(l_A, b_A)$ is the position of the Solar apex in Galactic coordinates.

**b) The solution from radial velocities**

Suppose we have chosen $(N_i)$ stars uniformly distributed over the sky. For each star, we have an equation of the heliocentric radial velocity ($V_r$; km s$^{-1}$) for the i$^{th}$ star,

$$V_r = V_x \cos\delta \cos\alpha + V_y \cos\alpha \sin\alpha + V_z \sin\alpha. \tag{14}$$

By definition, the Solar motion is determined relative to the centroid of the observed velocities, Equation (14) can be considered as the condition equation for the least-square solution $(V_x, V_y, V_z; \text{km s}^{-1})$, then the Sun's velocities concerning this same group and referred along x, y, and z-axes in the coordinate system whose centered at the Sun are given as; $(X_\odot^\bullet = -\bar{V}_X)$, $(Y_\odot^\bullet = -\bar{V}_Y)$, and $(Z_\odot^\bullet = -\bar{V}_Z)$. Therefore, we can calculate the absolute value of the Sun's velocity relative to our target and the position of the Solar apex in equatorial coordinates as follows.

$$S_\odot = \sqrt{(X_\odot^\bullet)^2 + (Y_\odot^\bullet)^2 + (Z_\odot^\bullet)^2}, \tag{15}$$

$$\alpha_A = \tan^{-1}\left(\frac{Y_\odot^\bullet}{X_\odot^\bullet}\right), \tag{16}$$

$$\delta_A = \tan^{-1}\left(\frac{Z_\odot^\bullet}{\sqrt{(X_\odot^\bullet)^2 + (Y_\odot^\bullet)^2}}\right), \tag{17}$$

where $(\alpha_A, \delta_A)$ is the position of the Solar apex in the equatorial coordinates.

Following the above computational scheme, a Mathematica routine has been developed to compute these kinematics and velocity ellipsoid parameters (VEPs) following the computational algorithm developed by Elsanhoury et al. (2015), Elsanhoury (2021), and Bisht et al. (2020). The numerical results are listed in Tables (1, 2, and 3) for different kinematical parameters like; velocity dispersion $(\sigma_1, \sigma_2, \sigma_3; \text{km s}^{-1})$, the ratio $(\sigma_2/\sigma_1)$, the longitude of the vertex $(l_2)$, and the Solar elements $(S_\odot, l_A, b_A, \alpha_A, \delta_A)$. We bold attention here on the self-comparison between these two sets of programs I, II, and III stars. The velocity dispersion $(\sigma_1, \sigma_2, \sigma_3; \text{km s}^{-1})$, $(\sigma_2/\sigma_1)$, and the Solar elements for these samples of stars and subsamples are quite comparable. Since one axis of the velocity ellipsoid is arranged normally to the Galactic plane so that the other two axes lie in the plane, therefore the longest axis of the ellipsoid points approximately to the Galactic center. To specify the orientation of the velocity ellipsoid, we need to determine only the Galactic longitude along which the principal axis lies, i.e., the longitude of the vertex $(l_2)$ (Mihalas & Binney, 1981), our results indicated that the longitude of the vertex $(l_2)$ often differs significantly from zero (e.g., Elsanhoury et al. 2015; Elsanhoury 2016; Bisht et al. 2020).

In Table (4) we compared the results from the present work for the total speed of the Solar motion ($S_\odot$; km s$^{-1}$), the ratio ($\sigma_2/\sigma_1$), and the ratio ($\sigma_3/\sigma_1$) calculated in the present work and that from other studies. The values of the velocity dispersions ($\sigma_1, \sigma_2, \sigma_3$) for the three program stars obey the inequalities $\sigma_1 > \sigma_2 > \sigma_3$ and behave in a radially elongated velocity ellipsoid. The total speed of the Solar motion $S_\odot$ is in good agreement with those of Ratnatunga & Upgren (1997) and Wehlau (1957) for M-dwarfs, but the ratios ($\sigma_2/\sigma_1$) and ($\sigma_3/\sigma_1$) for our three program stars differ slightly from that calculated for M-dwarfs.

**TABLE 1:** Kinematical parameters of the program I star retrieved from Newton et al. (2016).

| Parameters | Program I star, within Solar neighborhood ≈ 25 pc, 538 points | | |
|---|---|---|---|
| | All data | $b \geq 20º$ | $b < 20º$ |
| $N_i$ | 538 | 225 | 313 |
| $(\overline{U}, \overline{V}, \overline{W})$ (km/s) | -12.28, -24.14, -6.52 | -12.88, -26.84, -2.97 | -11.85, -22.20, -9.07 |
| $(\overline{V}_x, \overline{V}_y, \overline{V}_z)$ (km/s) | -5.37, 22.80, -15.08 | -9.75, 23.84, -15.23 | -2.23, 22.07, -14.97 |
| $(\lambda_1, \lambda_2, \lambda_3)$ (km/s) | 1929.72, 1069.29, 451.34 | 2144.65, 1149.28, 495.90 | 1783.41, 1011.71, 408.533 |
| $(\sigma_1, \sigma_2, \sigma_3)$ (km/s) | 43.93, 32.70, 21.25 | 46.31, 33.91, 22.27 | 42.23, 31.81, 20.21 |
| $(\sigma_2 / \sigma_1)$ | 0.75 | 0.73 | 0.75 |
| $(l_1, m_1, n_1)º$ | 0.8414, 0.5302, 0.1052 | 0.8421, 0.5380, 0.0380 | 0.8371, 0.5223, 0.1630 |
| $(l_2, m_2, n_2)º$ | -0.5410, 0.8258, 0.1608 | -0.5379, 0.8325, 0.1328 | -0.5457, 0.8182, 0.1809 |
| $(l_3, m_3, n_3)º$ | 0.0017, 0.1922, -0.9814 | 0.0398, -0.1323, 0.9904 | 0.0389, 0.2404, -0.9700 |
| $S_\odot$ (km/s) | 27.85 ± 5.28 | 29.92 ± 5.47 | 26.74 ± 5.17 |
| $(l_A, b_A)º$ | -63.04, 13.53 | -64.36, 5.69 | -61.91, 19.82 |
| $(\alpha_A, \delta_A)º$ | -76.75, 32.74 | -67.76, 30.36 | -84.23, 34.05 |

**TABLE 2:** Kinematical parameters of the program II stars retrieved from Newton et al. (2016)

| Parameters | Program II stars, located above 25 pc till ≈ 50 pc, 100 points | | |
|---|---|---|---|
| | All data | $b \geq 20º$ | $b < 20º$ |
| $N_i$ | 100 | 41 | 59 |
| $(\overline{U}, \overline{V}, \overline{W})$ (km/s) | -11.07, -22.65, -10.84 | -9.02, -24.29, -7.67 | -12.50, -21.51, -13.04 |
| $(\overline{V}_x, \overline{V}_y, \overline{V}_z)$ (km/s) | -0.94, 21.94, -16.47 | -4.61, 20.26, -17.29 | 1.61, 23.10, -15.90 |
| $(\lambda_1, \lambda_2, \lambda_3)$ (km/s) | 1768.14, 877.09, 432.15 | 1734.90, 911.47, 305.96 | 1813.42, 853.02, 484.85 |
| $(\sigma_1, \sigma_2, \sigma_3)$ (km/s) | 42.05, 29.62, 20.79 | 41.65, 30.20, 17.50 | 42.59, 29.21, 22.02 |
| $(\sigma_2 / \sigma_1)$ | 0.70 | 0.73 | 0.69 |
| $(l_1, m_1, n_1)º$ | 0.8607, 0.4491, 0.2399 | 0.8505, 0.5124, 0.1191 | 0.8549, 0.4033, 0.3264 |
| $(l_2, m_2, n_2)º$ | -0.4937, 0.8513, 0.1777 | -0.5237, 0.8459, 0.1012 | -0.4936, 0.8260, 0.2722 |
| $(l_3, m_3, n_3)º$ | 0.1244, 0.2713, -0.9544 | 0.04884, 0.1484, -0.9878 | 0.1600, 0.3938, -0.9052 |
| $S_\odot$ (km/s) | 27.44 ± 5.24 | 27.02 ± 5.20 | 28.08 ± 5.30 |
| $(l_A, b_A)º$ | -63.96, 23.27 | -69.65, 16.50 | -59.84, 27.66 |
| $(\alpha_A, \delta_A)º$ | -87.55, 36.98 | -77.18, 39.77 | 86.02, 34.65 |

**TABLE 3:** Kinematical parameters of the program III stars retrieved from Zhang et al. (2019).

| Parameters | Program III stars, located above 25 pc till ≈ 0.70 kpc, 60 points | | |
|---|---|---|---|
| | All data | $b \geq 20°$ | $b < 20°$ |
| $N_i$ | 60 | 26 | 34 |
| $(\bar{U}, \bar{V}, \bar{W})$ (km/s) | 3.32, -20.24, -7.22 | -6.22, -24.92, -7.87 | 10.61, -16.66, -6.72 |
| $(\bar{V}_x, \bar{V}_y, \bar{V}_z)$ (km/s) | -3.68, 7.60, -20.04 | -4.89, 18.13, -19.21 | -2.76, -0.47, -20.67 |
| $(\lambda_1, \lambda_2, \lambda_3)$ (km/s) | 1402.34, 854.30, 402.25 | 2158.72, 944.01, 565.88 | 1053.38, 626.91, 186.35 |
| $(\sigma_1, \sigma_2, \sigma_3)$ (km/s) | 37.45, 29.23, 20.06 | 46.46, 30.73, 23.79 | 32.46, 25.04, 13.65 |
| $(\sigma_2 / \sigma_1)$ | 0.78 | 0.67 | 0.77 |
| $(l_1, m_1, n_1)°$ | 0.3140, -0.9364, -0.1569 | 0.1203, -0.9924, -0.0266 | 0.7889, -0.5403, -0.2927 |
| $(l_2, m_2, n_2)°$ | -0.9495, -0.3095, -0.0525 | -0.9454, -0.1063, -0.3084 | -0.6103, -0.7447, -0.2701 |
| $(l_3, m_3, n_3)°$ | 0.0007, 0.1654, -0.9863 | 0.3032, 0.0623, -0.9509 | 0.0720, -0.3917, 0.9173 |
| $S_\odot$ (km/s) | 21.74 ± 4.66 | 26.86 ± 5.18 | 20.86 ± 2.28 |
| $(l_A, b_A)°$ | 80.70, 19.40 | -75.99, 17.04 | 57.51, 18.80 |
| $(\alpha_A, \delta_A)°$ | -64.17, 67.17 | -74.91, 45.65 | 9.66, 82.29 |

**TABLE 4:** The Solar velocity and velocity dispersion ratios for different spectral types.

| Spectral Types | $S_\odot (km\ s^{-1})$ | $(\sigma_2/\sigma_1)$ | $(\sigma_3/\sigma_1)$ | Reference |
|---|---|---|---|---|
| $\geq$ M3, ≈ 25 pc | 27.85 ± 5.28 | 0.74 | 0.48 | Present work, 538 points (program I) |
| $\geq$ M3, 25 pc till ≈ 50 pc | 27.44 ± 5.24 | 0.70 | 0.50 | Present work, 100 points (program II) |
| $\geq$ M3, 25 pc till ≈ 0.70 kpc | 21.74 ± 4.66 | 0.78 | 0.54 | Present work, 60 points (program III) |
| Hot WD | 11.28 | 0.60 | 0.80 | Elsanhoury et al. (2015) |
| cold WD | 21.97 | 0.72 | 0.85 | Elsanhoury et al. (2015) |
| DA WD | 18.78 | 0.69 | 0.83 | Elsanhoury et al. (2015) |
| Non-DA WD | 23.23 | 0.76 | 0.91 | Elsanhoury et al. (2015) |
| Cepheids kinematics, DR2 Catalog | 18.33 | - | - | Bobylev et al. (2021) |
| LAMOSTA-type Stars ≈ 0.70 kpc | 17.28 | 0.55 | 0.37 | Wang et al. (2021) |
| G, K, and M stars | 18.29 ± 0.19 | - | - | Branham (2017) |
| M – dwarfs ≈ 2.00 kpc | - | - | 0.60 | Binney et al. (2014) |
| G III | 16.72 ± 1.05 | 0.83 ± 0.06 | 0.55 ± 0.04 | Branham (2011) |
| M III | 24.21 ± 0.70 | 0.80 ± 0.09 | 0.59 ± 0.06 | Branham (2008) |
| O – B5 giants | 13.83 ± 0.17 | 0.81 ± 0.05 | 0.58 ± 0.05 | Branham (2006) |
| M – dwarfs ≈ 50 pc | 23.00 ± 3.6 | 0.67 | 0.58 | Ratnatunga & Upgren (1997) |
| < 8.0 $M_V$ | - | 0.82 ± 0.05 | 0.68 ± 0.04 | Dyer & Edward (1956) |
| 8.9 ≥ $M_V$ ≥ 8 | - | 0.72± 0.04 | 0.62 ± 0.04 | Dyer & Edward (1956) |
| $M_V$ ≥ 9.0 | - | 0.67± 0.05 | 0.56 ± 0.04 | Dyer & Edward (1956) |
| K8-M2 | 17.30 | 0.65 | 0.59 | Wehlau (1957) |
| M3-M6 | 23.30 | 0.74 | 0.52 | Wehlau (1957) |

## 4. THE OORT CONSTANTS

The orbital parameters of the Sun, such as the orbital velocity and period, and the local features of the Galactic disc such as the mass density and the changing of the rotating speed as a function of the radius from the Galactic center can be derived from the Oort constants. The Oort constants are connected to circular velocity and so the potential of the Galaxy in an axisymmetric approximation (Binney & Merrifield 1998).

The first-ever proof of the presence of the differential Galactic rotation was by Oort (1927a, 1927b). And since time, there are many formulas for the two Oort's constants (A) and (B). To evaluate the rotation constant (A), we obey the fact that the heliocentric radial velocity ($V_r$; km s$^{-1}$) shows a double sine-wave variance with the Galactic longitude with an amplitude that increases linearly with distance (Balona & Feast 1974).

$$V_r = -2A(R - R_o) \sin l \cos b + K, \tag{18}$$

where ($l$) and ($b$) are the longitude and latitude of the individual star, ($R_o$) is the distance from the Sun to the Galactic center and (K) is a term that can be interpreted as systematic motions of large stellar groupings, systematic errors in the radial velocities due to such causes as gravitational redshift, motions within stellar atmospheres, and erroneous wave-length systems (Feast & Shuttleworth 1965). The radial distance of the star from the Galactic center (R) (the cylindrical radius vector) is given by.

$$R^2 = R_o^2 + d^2 - 2R_o\, d \cos l. \tag{19}$$

We calculated the Oort constants (A) and (B) for the three program stars (five subsamples) as listed in Table (5), namely: Program I (538 stars), Program II (100 stars), and Program III (60 stars). In Table (5), Column 2 is the ratio ($\sigma_2/\sigma_1$) computed by the computational method described in section 3, column 3 is the first Oort constant (A; km s$^{-1}$kpc$^{-1}$) computed from the least-squares fit of Equation (18), column 4 is the second Oort constant (B; km s$^{-1}$kpc$^{-1}$) computed using the relation $(\sigma_2/\sigma_1)^2 = -B/(A - B)$ (Mihalas & Binney 1981), column 5 is the (K $-$ term; km s$^{-1}$), column 6 represent the angular velocity ($|A - B|$; km s$^{-1}$kpc$^{-1}$), and in column 7 we listed the rotational velocity ($V_o$; km s$^{-1}$) computed following the well-known relation $V_o = |A - B|R_o$. In the present calculations, we used $R_o = 8.20 \pm 0.10$ kpc as recently calculated by Bland-Hawthorn et al. (2019).

In Table (6), we listed Oort constants adopted by some earlier studies. Bovy (2017) used data for 304,267 main sequence stars from the Gaia DR1 to calculate Oort constants at a typical heliocentric distance of 230 pc. Based on the Gaia DR2 data, Li et al. (2019) calculated Oort constants and derive the Galactic rotational properties for a sample of stars within 500 pc. Krisanova et al. (2020) calculated Oort constants using proper motions and trigonometric parallaxes of more than 25,000 young stars from the Gaia DR2 catalog. Nouh & Elsanhoury (2020) estimated Oort constants using a sample of halo red giants based on the space and radial velocities of 1,583 red giant stars collected from the SEGUE-1 and SEGUE-2 surveys. Wang et al. (2021) derived Oort constants using a sample of 5,627 A-type stars within 0.60 kpc selected from the LAMOST surveys.

Comparing the results from the present analysis with that listed in Table (6) revealed good agreement for the first Oort constant (A), but we have marked different values for the second Oort constant (B). This difference may be attributed to the different methods of calculations. The Oort constants listed in Table (6) are calculated using proper motions except Nouh & Elsanhoury (2020), while in the present paper we calculated (A) by fitting the radial velocities to Equation (18), and (B) is computed using the relation $(\sigma_2/\sigma_1)^2 = -B/(A - B)$. The overestimated value of (B) leads to a notable difference between the present estimated value of $|A - B|$ and that from other studies.

**TABLE 5:** Rotation constants for the five samples under study.

| Sample | $(\sigma_2/\sigma_1)$ | A km s$^{-1}$ kpc$^{-1}$ | B km s$^{-1}$ kpc$^{-1}$ | K − term km s$^{-1}$ | \|A − B\| km s$^{-1}$ kpc$^{-1}$ | V$_o$ km s$^{-1}$ |
|---|---|---|---|---|---|---|
| Program I (538 stars) | 0.74 | 16.055±0.68 | -19.433±1.37 | 3.37 | 35.500 | 291.100±16.86 |
| Program I (225 stars) | 0.73 | 15.180±0.12 | -17.318±0.31 | 2.13 | 32.498 | 266.483±8.54 |
| Program I (313 stars) | 0.75 | 11.595±1.54 | -14.905±0.89 | 3.31 | 26.500 | 217.300±20.18 |
| Program II (100 stars) | 0.70 | 16.000±0.65 | -15.140±0.77 | -0.86 | 31.140 | 255.348±1.01 |
| Program III (60 stars) | 0.73 | 14.619±0.03 | -16.678±0.01 | 2.06 | 31.297 | 256.635±0.37 |

**TABLE 6:** Rotation constants from different studies

| A km s$^{-1}$ kpc$^{-1}$ | B km s$^{-1}$ kpc$^{-1}$ | \|A − B\| km s-1 kpc-1 | Method | Reference |
|---|---|---|---|---|
| $15.30 \pm 0.40$ | $-11.90 \pm 0.40$ | 27.20 | Proper motions | Bovy (2017) |
| $15.10 \pm 0.10$ | $-13.40 \pm 0.40$ | 28.40 | Proper motions | Li et al. (2019) |
| $15.73 \pm 0.32$ | $-12.67 \pm 0.34$ | 28.40 | Proper motions | Krisanova et al. (2020) |
| $15.60 \pm 1.60$ | $-13.90 \pm 1.80$ | 29.50 | Radial velocities | Nouh & Elsanhoury (2020) |
| $16.31 \pm 0.89$ | $-11.99 \pm 0.79$ | 28.30 | Proper motions | Wang et al. (2021) |
| $14.69 \pm 0.61$ | $-16.70 \pm 0.67$ | 31.39 | Radial velocities | Present work |

## 5. DISCUSSION AND CONCLUSIONS

In the present paper, we calculated the kinematical parameters and the Oort constants (A) and (B) for three program stars of M-dwarf stars. Besides the three main samples, we divided program I into two subsamples according to the Galactic latitudes namely for high latitude $b \geq 20º$ and low one (i.e., $b < 20º$). For each sample, the velocities, the projected distances, and the Solar elements are calculated and listed as shown in Tables (1, 2, and 3).

Dwarf stars velocity ellipsoid dimensions vary depending on the history of their dynamical excitation by the molecular clouds and spiral arms (Binney & Tremaine 2008), for hot dwarfs (T > 12,000 K), we have velocity ellipsoidal components $(\sigma_1, \sigma_2, \sigma_3, S_\odot; \text{km s}^{-1}) = (29.10, 17.50, 23.31, 11.28; \text{km s}^{-1})$ (see Elsanhoury et al. 2015) inequalities for a cold one (e.g., M – dwarfs; T < 12,000 K) and the velocity ellipsoid appears radially longer, e.g., $(\sigma_1, \sigma_2, \sigma_3, S_\odot; \text{km s}^{-1}) = (43.93, 32.70, 21.25, 27.85 \pm 5.28; \text{km s}^{-1})$ as listed in Table (4) with our present work, which could be devoted for both types to different in sizes (Rafikov 2018) and their turbulent motion (Hunter & Flecker 1982; Flicker 1982).

The longitude of the vertex $(l_2)$ often differs significantly from zero and is affected with stellar spectral classes (i.e., temperature). The obtained numerical values of the longitude of the vertex $(l_2)$ are: $-0°.5410$ (538 points), $-0°.4937$ (100 points), and $-0°.9495$ (60 points).

The calculated Oort constants listed in Table (5) show consistent values except for program I (313 stars), $A = 11.595 \pm 1.54 \text{ km s}^{-1}\text{kpc}^{-1}$. We adopt the mean value of the Oort constants as $A = 14.68 \pm 0.61 \text{ km s}^{-1}\text{kpc}^{-1}$ and $B = -16.70 \pm 0.67 \text{ km s}^{-1}\text{kpc}^{-1}$. The rotational velocity $(V_o; \text{km s}^{-1})$ listed in Table (5) ranged from $217 - 290 \text{ km s}^{-1}$ and we compute $|A - B| = 31.39 \text{ km s}^{-1}\text{kpc}^{-1}$ and $V_o = 257.38 \pm 9.40 \text{ km s}^{-1}$.

There are several values for the rotational velocity calculated by earlier investigations. For example, Oort (1927a) and (1927b) adopt it as $247.50 \text{ km s}^{-1}$, a direct measurement of $(V_o; \text{km s}^{-1})$; based on radial velocities of globular clusters or spheroidal-component stars in our Galaxy or external Galaxies in the Local Group, yield in the range $(200 \leq V_o \leq 300 \text{ km s}^{-1})$ (Mihalas & Binney 1981); $204 \text{ km s}^{-1}$ (Bovy 2017), $213.7 \text{ km s}^{-1}$ (Li et al. 2019), and recently we determined $V_o = 221.25 \pm 26.66 \text{ km s}^{-1}$ (Nouh & Elsanhoury 2020). The effect of using proper motions of faint stars and that have large distances from the Sun on the calculations of the kinematical parameters is demonstrated by Lewis (1990). This effect gives an underestimate or overestimates of the value of the Oort constants (A) and (B). This may explain the overestimated values of the second Oort constant (B) as it is calculated using the relation $(\sigma_2/\sigma_1)^2 = -B/(A - B)$. The difference between our rotational velocity (~257 km/s) and Gaia value (~232 km/s); assuming that $R_o = 8.20 \pm 0.10$ kpc; may be attributed to the different samples used in the two analyses.

## ACKNOWLEDGMENTS

The authors thank the anonymous referee for the useful comments that improved the scientific content of the article significantly. This work presents results from the European Space Agency (ESA) space mission, Gaia. Gaia data are being processed by the Gaia Data Processing and Analysis Consortium (DPAC). Funding for the DPAC is provided by national institutions, in particular the institutions participating in the Gaia Multi-Lateral Agreement (MLA). The Gaia mission website is https://www.cosmos.esa.int/gaiaall. The Gaia archive website is https://archives.esac.esa.int/gaia.

## CONFLICT OF INTEREST

The authors declare no financial or commercial conflict of interest.

## REFERENCES

Balona, L. A. and Feast, M. W., 1974, MNRAS, 167, 621.

Berta, Z. K., Irwin, J., Charbonneau, D., Burke, C. J., and Falco, E. E., 2012, AJ, 144, 145.

Binney, J., and Merrifield, M., 1998, Galactic Astronomy/James Binney and Michael Merrifield (Princeton, NJ: Princeton Univ. Press).

Binney, J., and Tremaine, S., 2008, Galactic Dynamics; Second Edition (Princeton University Press).

Binney, J., Burnett, B., Kordopatis, G., Steinmetz, M., Gilmore, G., Bienayme, O., Bland-Hawthorn, J., Famaey, B., Grebel, E. K., Helmi, A., Navarro, J., Parker, Q., Reid, W. A., Seabroke, G., Siebert, A., Watson, F., Williams, M. E. K., and Wyse, R. F. G., Zwitter, T., 2014, MNRAS, 439, 1231.

Bisht, D., Elsanhoury, W. H., Zhu, Q., Sriya, D. P. et al., 2020, AJ, 160, 119.

Bland-Hawthorn, J., Sharma, S. et al., 2019, MNRAS, 486, 1167.

Bobylev, V. V., Bajkova, A. T., Rastorguev, A. S., and Zabolotskikh, M. V., 2021, MNRAS, 502, 4377.

Bochanski, J. J., Savcheva, A., and West, A. A., 2013, AJ, 145, 40.

Bovy J., 2017, MNRAS, 468, L63.

Branham, R. L. Jr., 20018, Rev. Mex. Astron. Astrofís., 44, 29.

Branham, R. L. Jr., 2006, MNRAS, 370, 1393.

Branham, R. L. Jr., 2011, MNRAS, 47, 197.

Branham, R. L. Jr., 2014, Astrophys. Space Sci., 353, 179.

Branham, R. L. Jr., 2017, Astrophys. Space Sci., 362, 29.

Chabrier, G., and Baraffe, I., 1997, A&A, Vol. 327, 1039.

Charbonneau, D., Berta, Z. K., Irwin, J., Burke, C. J., Nutzman, P., Buchhave, L. A., Lovis, C., Bonfils, X., et al., 2009, Nature, 462, 891.

Cui, X.-Q., Zhao, Y.-H., Chu, Y.-Q., et al., 2012, RAA, 12, 1197

Delfosse, X., Forveille, T., Perrier, C., and Mayor, M., 1998, A&A, 331, 58.

Dyer, Edward R., Jr., 1996, AJ., 61, 228.

Elisabeth R. Newton, Jonathan Irwin, David Charbonneau, Zachory K. Berta Thompson, Jason A. Dittmann1, and Andrew A. West, 2016, ApJ, 821, 93.

Elsanhoury, W. H., 2016, Ap, 59, pp. 246.

Elsanhoury, W. H., 2021, JOAA, in press.

Elsanhoury, W. H., Nouh, M. I. and Abdel-Rahman, H. I., 2015, Rev. Mex. Astron. Astrofís., 51, 197.

Feast, M. W.; Shuttleworth, M., 1965, MNRAS, 130, 243.

Flecker, R. C., 1982, MNRAS, 201, 551.

Gaia Collaboration, Brown, A. G. A., Vallenari, A., et al., 2018, A&A, 616, A1

Gaia Collaboration, Brown, A. G. A., Vallenari, A., et al., 2020, A&A, 649, A1.

Gaidos, E., Haghighipour, N., Agol, E., Latham, D., Raymond, S., and Rayner, J. 2007, Science, 318, 210

Guinan, E. F., Engle, S. G., and Durbin A., 2016, ApJ, 821, 81.

Habson, M. J., Jofre, E, Garcia, L., Petrucci, R., and Gomez, M., Rev. Mex. Astron. Astrofís., 2018, 54, 65–84.

Hunter, J. H., and Flecker, R. C., 1982, ApJ, 256, 505.

Irwin, J. M., Berta-Thompson, Z. K., Charbonneau, D., et al., 2014, in Proc. Conf. 767, 18th Cambridge Workshop on Cool Stars, Stellar Systems, and the Sun, ed. G. van Belle, and H. C. Harris.

Jaschek, C. and Valbousquet, A., 1991, A&A, 242, 77.

Jaschek, C. and Valbousquet, A., 1992, A&A, 252, 124.

Jaschek, C. and Valbousquet, A., 1993, A&A, 275, 472.

Jaschek, C. and Valbousquet, A., 1994, A&A, 291, 448.

Kaltenegger, L. and Traub, W. A. 2009, ApJ, 698, 519.

Kirkpatrick, J. D., 2005, ARA&A, 43, 195.

Kirkpatrick, J. D., Henry, T. J., and McCarthy, D. W. J., 1991, ApJSS, 77, 417.

Kroupa, P., Weidner, C., Pflamm-Altenburg, J., et al., 2013, The Stellar and Sub-Stellar Initial Mass Function of Simple and Composite Populations, (Dordrecht: Springer).

Krisanova, O. I., Bobylev, V. V., and Bajkova. A. T., 2020, Astron. Lett., 46, 6.

Kuiper, G. P., 1939, ApJ, 89, 548

Laughlin, G., and Adams, F. C., 1997, ApJ, 491, 51.

Lewis, J. R., 1990, MNRAS, 244, 247.

Li, C., Zhao, G., and Yang, C., 2019, ApJ, 872, 205.

Lindegren, L., Kiloner, S. A., Hernández, J., et al., 2021, A&A, 649, A2 (Gaia EDR3 SI).

Liu, J.-C., Zhao, Z., and Hu, B., 2011, A&A, 536, A102.

Lunine, J. I., Macintosh, B., and Peale, S., 2009, Physics Today, 62, 050000

Martín, E. L., Delfosse, X., Basri, G., Goldman, B., Forveille, T., and Zapatero Osorio, M. R., 1999, AJ, 118, 2466.

Mihalas, D. and Binney, J., 1981, Galactic Structure "Structure and Kinematics", $2^{nd.}$ Edition W. H. Freeman and Company – San Francisco.

Mihalas, D., and Binney, J., 1968, Galactic Astronomy ($2^{nd.}$ Edition; New York: Freeman).

Mohanty, S., and Basri, G., 2003, ApJ, 583, 451.

Muirhead, P. S., Edelstein, J., Erskine, D. J., et al., 2011, PASP, 123, 709.

Nouh, M. I. and Elsanhoury, W. H., 2020, Ap, Vol. 36, No. 2, pp. 179-189.

Nutzman, P. and Charbonneau, D., 2008, PASP, 120, 317.

Oort, J. H., 1927a, Bull. Astron. Inst. Netherl, 3, 275.

Oort, J. H., 1927b, Bull. Astron. Inst. Netherl., 4, 91.

Rafikov, R. R., 2018, ApJ, 861, 35.

Ratnatunga, K. U. and Upgren, A. R., 1997, AJ, 476, 811.

Reiners, A., and Basri, G., 2007, ApJ, 656, 1121.

Smart, W. M., 1968, Stellar Kinematics (London: Longmans).

Taylor, M. B. 2005, in Astronomical Data Analysis Software and Systems XIV, eds. P. Shopbell, M. Britton, and R. Ebert, ASP Conf. Ser., 347, 29.

Torra, F., Castañede, J., Fabricius, C., et al., 2020, arXiv:2012.06420.

Wang, F., Zhang, H. -W., Huang, Y., Chen, B. -Q., Wang, H. -F., Wang, C., 2021, MNRAS, 504, 199.

Wehlau, Amelia White, 1957, AJ, 62, 169.

Weiler, M. 2018, A&A, 617, A138.

West, A. A., Hawley, S. L., Bochanski, J. J., Covey, K. R., Reid, I. N., Dhital, S., Hilton, E. J., and Masuda, M., 2008, AJ, 135, 785.

Zhang, Shuo; Luo, A. -Li; Comte, Georges; Gizis, John E.; Wang, Rui; Li, Yinbi; Qin, Li; Kong, Xiao; Bai, Yu; Yi, Zhenping, ApJS., 2019, Vol. 240, 31, pp. 15.